\documentclass[amsmath,10pt]{wlscirep}
\usepackage[utf8]{inputenc}
\usepackage[T1]{fontenc}
\usepackage{txfonts}
\title{Quantum entanglement maintained by virtual excitations in an ultrastrongly-coupled-oscillator system}
\author{Jian-Yong Zhou}
\author{Yue-Hui Zhou}
\author{Xian-Li Yin}
\author[1,*]{Jin-Feng Huang}
\author[1,*]{Jie-Qiao Liao}
\affil{Department of Physics and Synergetic Innovation Center for Quantum Effects and Applications, Hunan Normal University, Changsha 410081, China. Key Laboratory of Low-Dimensional Quantum Structures and Quantum Control of Ministry of Education, Hunan Normal University, Changsha 410081, China. Key Laboratory for Matter Microstructure and Function of Hunan Province, Hunan Normal University, Changsha 410081, China. }
\affil[*]{jfhuang@hunnu.edu.cn; jqliao@hunnu.edu.cn}

\begin{abstract}
We study the effect of quantum entanglement maintained by virtual excitations in an ultrastrongly-coupled harmonic-oscillator system. Here, the quantum entanglement is caused by the counterrotating interaction terms and hence it is maintained by the virtual excitations. We obtain the analytical expression for the ground state of the system and analyze the relationship between the average excitation numbers and the ground-state entanglement. We also study the entanglement dynamics between the two oscillators in both the closed- and open-system cases. In the latter case, the quantum master equation is microscopically derived in the normal-mode representation of the coupled-oscillator system. This work will open a route to the study of quantum information processing and quantum physics based on virtual excitations.
\end{abstract}
\begin{document}

\flushbottom
\maketitle
\thispagestyle{empty}

\section*{Introduction}
The ultrastrong coupling (USC) physics~\cite{SolanoRMP,Kockum2019} has recently attracted much attention from the communities of quantum physics, quantum optics, and condensed matter physics. Great advances have been made in both theory~\cite{Ciuti2005,Liberato2007,Bourassa2009,Anappara2009,Casanova2010
,Ashhab2010,Beaudoin2011,Ballester2012,Huang2014,Pedernales2015,Kockum2017pra,Garziano2016prl,Stassi2015pra} 
and experiments in various physical platforms, including semiconductor cavity quantum electrodynamical (QED) systems~\cite{Huber2009Nat,Sirtori2010PRL,Faist2012}, superconducting circuit-QED systems~\cite{Gross2010NatPhy,Mooij2010PRL,Semba2017,Lupascu2017,You2017,Steele2017}, coupled photon-2D-electron-gas~\cite{Faist2012Sci,Wegscheider2014,Kono2016}, light-molecule~\cite{Ebbesen2011,Gigli2014}, and photon-magnon systems~\cite{Tobar2014}.
In the USC regime~\cite{SolanoRMP,Kockum2019}, the coupling strength is comparable to the transition frequencies in the system, and then the rotating-wave approximation (RWA) is invalid, namely the counterrotating (CR) terms should be kept in the interactions. It has been demonstrated that the CR terms could produce some unpredictable physical phenomena~\cite{Ciuti2005} and have wide applications in quantum information processing~\cite{Felicetti2014,Romero2012}. In particular, the development of the ultrastrong coupling field promotes various studies in quantum optics topics beyond the RWA such as the quantum Rabi model~\cite{Braak2011PRL,Chen2012PRA,Lee2013JPA,Xie2014PRX,Huang2015,Klimov2009book,Zheng2008PRL}.

One of the interesting effects associated with the CR terms in the USC regime is the generation of virtual excitations. In the presence of the CR terms, the ground states of some typical quantum systems possess virtual excitations. For example, in the quantum Rabi model, it has been shown that virtual photons exist in the ground state~\cite{Huang2014}. These virtual photons cannot be detected directly even if this absorber is placed inside the cavity, except with very small probability at short times set by the time-energy uncertainty~\cite{Stefano2018}. On the basis of these properties, the ground-state photons in the USC system are considered virtual photons~\cite{Kockum2019}. However, even though these virtual photons cannot be detected directly, there are still ways to probe them. One proposal is to measure the change that they produce in the Lamb shift of an ancillary probe qubit coupled to the cavity~\cite{Ciuti2015}. Another proposal is to detect the radiation pressure that they give rise to if the cavity is an optomechanical system~\cite{Nori2017}. These proposals rely on the rapid modulation of either $g$ (light-matter coupling strength) or the atomic frequency. Then the virtual photons can be converted into real ones and extracted from the system~\cite{Beaudoin2011,Ciuti2005,Liberato2007,Liberato2009,Takashima2008,Werlang2008,Carusotto2012,Garziano2013,Shapiro2015}.

In this paper, we propose to study another quantum effect, quantum entanglement, associated with the virtual excitations.  Here the quantum entanglement is created by the CR terms and hence it is maintained by the virtual excitations.
We note that the relationship between quantum entanglement and the CR terms has been previously considered in the quantum Rabi model~\cite{Ashhab2010}. In addition, the role of the CR terms in the creation of entanglement between two atoms has been investigated in Ref.~\cite{Ficek2010}. We consider an ultrastrongly-coupled two-harmonic-oscillator system. We study the ground state entanglement of the two oscillators and analyze the average excitation numbers in the system. We also study the entanglement dynamics of the system when it is initially in the zero-excitation state and hence all the excitations are created by the CR terms. The influence of the environment dissipations on the system is analyzed based on a microscopically derived quantum master equation in the normal-mode representation.

The rest of this paper is organized as follows. Firstly, we present the physical model of two coupled harmonic oscillators and the Hamiltonian, we also analyze the property of the parity chain in this system. Secondly, we obtain the exact analytical eigensystem of the coupled two-oscillator system. Thirdly, the average virtual excitation numbers are calculated analytically and the quantum entanglement of the ground state is analyzed by calculating the logarithmic negativity. Fourthly, we study the dynamics of the average virtual excitation numbers and quantum entanglement between the two oscillators in both the closed- and open-system cases. Finally, we present a brief conclusion.
%%%%%%%%%%%%%%%%%%%%%%%%%
\begin{figure}[tbp]
\includegraphics[width=0.47\textwidth]{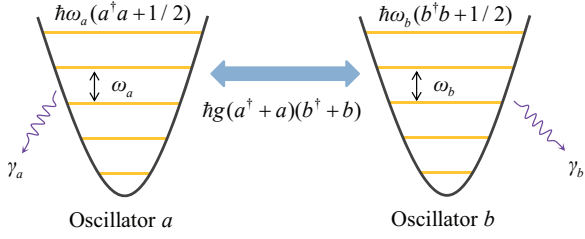}
\caption{Schematic diagram of the coupled two-harmonic-oscillator system. Two harmonic oscillators with resonance frequencies $\omega_{a}$ and $\omega_{b}$ are coupled to each other via a ``position-position" type interaction with the coupling strength $g$. The parameters $\gamma_{a}$ and $\gamma_{b}$ are the decay rates associated with the heat baths in contacted with the oscillators $a$ and $b$, respectively.}
\label{modelabcm}
\end{figure}
%%%%%%%%%%%%%%%%%%%%%%%%%

\section*{Results}
\textbf{Model and Hamiltonian.} We consider an ultrastrong coupling system, in which two harmonic oscillators are ultrastrongly coupled to each other through the so-called ``position-position" type interaction (Fig.~\ref{modelabcm}).
This system is described by the Hamiltonian
\begin{equation}
H=\sum_{i=1}^{2}\left(\frac{p_{i}^{2}}{2\mu}+\frac{\mu}{2}\omega_{i}^{2}x_{i}^{2}\right)+\eta(x_{1}-x_{2})^{2},\label{mhh}
\end{equation}
where $x_{1}$ ($x_{2}$) and $p_{1}$ ($p_{2}$) are, respectively, the coordinate and momentum operators of the first (second) oscillator with the resonance frequency $\omega_{1}$ ($\omega_{2}$) and mass $\mu$, the parameter $\eta$ is the coupling strength between the two oscillators.
By expanding the interaction term, Hamiltonian~(\ref{mhh}) can be expressed as
\begin{equation}
H=\sum_{i=1}^{2}\frac{p_{i}^{2}}{2\mu}+\frac{1}{2}\mu\omega_{a}^{2}x_{1}^{2}
+\frac{1}{2}\mu\omega_{b}^{2}x_{2}^{2}-\xi x_{1}x_{2},\label{mhh2}
\end{equation}
where we introduce the renormalized frequencies and coupling strength as
\begin{equation}
\omega_{a}=\sqrt{\omega_{1}^{2}+2\eta/\mu},\hspace{0.3 cm}
\omega_{b}=\sqrt{\omega_{2}^{2}+2\eta/\mu},\hspace{0.3 cm}
\xi=2\eta.
\end{equation}
By introducing the following creation and annihilation operators
\begin{equation}
a=(a^{\dag})^{\dag}=\sqrt{\mu\omega_{a}/(2\hbar)}x_{1}+i\sqrt{1/(2\mu\hbar\omega_{a})}p_{1},\hspace{0.3 cm}b=(b^{\dag})^{\dag}=\sqrt{\mu\omega_{b}/(2\hbar)}x_{2}+i\sqrt{1/(2\mu\hbar\omega_{b})}p_{2},
\end{equation}
Hamiltonian~(\ref{mhh2}) becomes
\begin{equation}
H=\hbar\omega_{a}a^{\dag}a+\hbar\omega_{b}b^{\dag}b+\hbar g(a^{\dag}+a)(b^{\dag}+b)+C,\label{eqHtwom}
\end{equation}
with $C=(\hbar\omega_{a}+\hbar\omega_{b})/2$ being a constant term. Here $a^{\dagger}$ ($a$) and $b^{\dagger}$ ($b$) are, respectively, the creation (annihilation) operators of the two oscillators with the corresponding resonance frequencies $\omega_{a}$ and $\omega_{b}$. In Eq.~(\ref{eqHtwom}), the first two terms and the constant term represent the free Hamiltonian of the two oscillators. The parameter $g=-\xi/(2\mu\sqrt{\omega_{a}\omega_{b}})$ denotes the coupling strength between the two oscillators. We note that this interaction includes both the rotating-wave and CR terms. In general, in the case of weak coupling and near resonance, the rotating-wave approximation can be made by discarding the CR terms. In this paper, we consider the ultrastrong-coupling case in which the CR terms cannot be discarded. In the presence of the CR terms, the ground state of the system will include excitations and hence quantum entanglement will exist in the ground state. Note that an ultrastrongly-coupled two-mode system has recently been realized in superconducting circuits~\cite{Teufel2019}.

In this two-oscillator system, we introduce the parity operator as $P=(-1)^{a^{\dag}a+b^{\dag}b}$, which has the standard properties of a parity operator, such as $P^{2}=I$, $P^{\dag}P=I$, and $P^{\dag}=P$~\cite{Casanova2010,Malekakhlagh2019}. The Hamiltonian $H$ in Eq.~(\ref{eqHtwom}) remains invariant under the transformation $P^{\dag}HP=H$, based on the relations $P^{\dag}aP=-a$, $P^{\dag}a^{\dag}P=-a^{\dag}$, $P^{\dag}bP=-b$, and $P^{\dag}b^{\dag}P=-b^{\dag}$. The Hilbert space of the system can be divided into two subspaces with different parities: odd and even. The basis states of the odd- and even-parity subspaces are, respectively, given by
\begin{eqnarray}
&&\{\vert 1,0\rangle_{a,b},\vert 0,1\rangle_{a,b}\}\leftrightarrow\{\vert 3,0\rangle_{a,b},\vert 2,1\rangle_{a,b},\vert 1,2\rangle_{a,b},\vert 0,3\rangle_{a,b}\}\nonumber\\
&&\leftrightarrow\{\vert 5,0\rangle_{a,b},\vert 4,1\rangle_{a,b},\vert 3,2\rangle_{a,b},\vert 2,3\rangle_{a,b},\vert 1,4\rangle_{a,b},\vert 0,5\rangle_{a,b}\}\leftrightarrow\cdots\nonumber\\
&&\leftrightarrow\{\vert 2n+1,0\rangle_{a,b},\cdots\vert n+1,n\rangle_{a,b},\vert n,n+1\rangle_{a,b},\cdots\vert 0,2n+1\rangle_{a,b}\}\leftrightarrow\cdots,
\end{eqnarray}
and
\begin{eqnarray}
&&\vert 0,0\rangle_{a,b}\leftrightarrow\{\vert 2,0\rangle_{a,b},\vert 1,1\rangle_{a,b},\vert 0,2\rangle_{a,b}\}\leftrightarrow\{\vert 4,0\rangle_{a,b},\vert 3,1\rangle_{a,b},\vert 2,2\rangle_{a,b},\vert 1,3\rangle_{a,b},\vert 0,4\rangle_{a,b}\}\leftrightarrow\cdots\nonumber\\
&&\leftrightarrow\{\vert 2n,0\rangle_{a,b},\cdots\vert n+1,n-1\rangle_{a,b},\vert n,n\rangle_{a,b},\vert n-1,n+1\rangle_{a,b},\cdots\vert 0,2n\rangle_{a,b}\}\leftrightarrow\cdots.
\end{eqnarray}
The eigenvalues of the parity operator $P$ corresponding to the odd and even parity states are $-1$ and $1$, respectively.

\textbf{Eigensystem of the coupled two-oscillator system.} To study the quantum entanglement of the eigenstates, we need to diagonalize the Hamiltonian $H$ in Eq.~(\ref{mhh2}). To this end, we introduce the transformation operator~\cite{Wagner1986}
\begin{equation}
U=\exp\left[-i\frac{\theta}{\hbar }\left(x_{1}p_{2}-x_{2}p_{1}\right)\right],
\end{equation}
where the mixing angle $\theta$ is defined by
\begin{equation}
\tan(2\theta)=\frac{2\xi}{\mu(\omega_{a}^{2}-\omega_{b}^{2})}.
\end{equation}
In terms of the transformation, the Hamiltonian in Eq.~(\ref{mhh2}) can be diagonalized as
\begin{eqnarray}
\tilde{H} =UHU^{\dag}=\frac{p_{1}^{2}}{2\mu}+\frac{\mu\omega _{A}^{2}x_{1}^{2}}{2}+\frac{p_{2}^{2}}{
2\mu}+\frac{\mu\omega _{B}^{2}x_{2}^{2}}{2},\label{Htildexp}
\end{eqnarray}
where the resonance frequencies are defined by
\begin{equation}
\omega_{A}^{2}=\frac{1}{2}\left(\omega_{a}^{2}+\omega_{b}^{2}\right)+
\frac{1}{2}\sqrt{\left(\omega_{a}^{2}-\omega_{b}^{2}\right)^{2}+4\xi^{2}/\mu^{2}},\hspace{0.3 cm}
\omega_{B}^{2}=\frac{1}{2}\left(\omega_{a}^{2}+\omega_{b}^{2}\right)-
\frac{1}{2}\sqrt{\left(\omega_{a}^{2}-\omega_{b}^{2}\right)^{2}+4\xi^{2}/\mu^{2}},
\end{equation}
with $\xi=-2\mu g\sqrt{\omega_{a}\omega_{b}}$.
By introducing the annihilation and creation operators
\begin{equation}
A=(A^{\dag})^{\dag}=\sqrt{\mu\omega _{A}/(2\hbar)}x_{1}+i\sqrt{1/(2\mu\hbar\omega_{A})}p_{1},\hspace{0.3 cm}
B=(B^{\dag})^{\dag}=\sqrt{\mu\omega_{B}/(2\hbar)}x_{2}+i\sqrt{1/(2\mu\hbar\omega_{B})}p_{2},
\end{equation}
Hamiltonian~(\ref{Htildexp}) can be expressed as
\begin{equation}
\tilde H=\hbar\omega_{A}A^{\dag}A+\hbar\omega_{B}B^{\dag}B+\frac{1}{2}\hbar(\omega_{A}+\omega_{B}).\label{HdiagAB2HO}
\end{equation}
The relations between the operators $A$ ($A^{\dag}$), $B$ ($B^{\dag}$), $a$ ($a^{\dag}$), and $b$ ($b^{\dag}$) can be obtained as
\begin{equation}
UaU^{\dag}=f_{1}A+f_{2}A^{\dag}+f_{3}B+f_{4}B^{\dag},\hspace{0.3 cm}
UbU^{\dag}=-f_{5}A-f_{6}A^{\dag}+f_{7}B+f_{8}B^{\dag}.
\end{equation}
Here the concrete forms of coefficients $f_{i}$ ($i=1,2,\cdots,7,8$) have been given by
\begin{subequations}
\begin{align}
f_{1,2}=&\frac{1}{2}\frac{\cos(\theta)}{\sqrt{\omega_{a}\omega_{A}}}(\omega_{a}\pm\omega _{A}),\hspace{0.3 cm}
f_{3,4}=\frac{1}{2}\frac{\sin(\theta)}{\sqrt{\omega_{a}\omega_{B}}}(\omega_{a}\pm\omega _{B}),\\
f_{5,6}=&\frac{1}{2}\frac{\sin(\theta)}{\sqrt{\omega_{b}\omega_{A}}}(\omega_{b}\pm\omega _{A}),\hspace{0.3 cm}
f_{7,8}=\frac{1}{2}\frac{\cos(\theta)}{\sqrt{\omega_{b}\omega_{B}}}(\omega_{b}\pm\omega_{B}).
\end{align}
\end{subequations}

Based on Eq.~(\ref{HdiagAB2HO}), we know the eigenstates of the system in the representation associated with $A^{\dag}A$ and $B^{\dag}B$ as
\begin{equation}
\tilde{H}\vert m\rangle_{A}\vert n\rangle_{B}=E_{m,n}\vert m\rangle_{A}\vert n\rangle_{B},\hspace{0.3 cm}m,n=0,1,2,\cdots,
\end{equation}
where the eigenvalues are given by
\begin{eqnarray}
E_{m,n}=\hbar\omega_{A}m+\hbar\omega_{B}n+\frac{1}{2}\hbar(\omega _{A}+\omega_{B}).\label{eigenvalueAB}
\end{eqnarray}
It is obvious that the ground state of the two-oscillator system is $\vert 0\rangle_{A}\vert 0\rangle_{B}$. To study the virtual excitations in the system, we need to know the eigenstates which are expressed in the representation associated with $a^{\dag}a$ and $b^{\dag} b$. It implies that we need to diagonalize the Hamiltonian $\tilde{H}$ in the representation of $a$ and $b$. To this end, we express Hamiltonian~(\ref{Htildexp}) with the bosonic creation (annihilation) operators $a^{\dag}$ $(a)$ and $b^{\dag}$ $(b)$ as
\begin{equation}
\tilde{H}=\frac{1}{2}\hbar(\omega_{A}^{2}/\omega_{a}+\omega_{a})({a}^{\dagger}{a}+1/2)+\hbar g_{a}({a}^{\dagger2}+{a}^{2})+\frac{1}{2}\hbar(\omega_{B}^{2}/\omega_{b}+\omega_{b})({b}^{\dagger}{b}+1/2)+\hbar g_{b}({b}^{\dagger2}+{b}^{2}),\label{habrep2HO}
\end{equation}
where we introduce the coupling strengths
\begin{equation}
g_{a}=\frac{1}{4}(\omega_{A}^{2}/\omega_{a}-\omega_{a}),\hspace{0.5 cm}
g_{b}=\frac{1}{4}(\omega_{B}^{2}/\omega_{b}-\omega_{b}).
\end{equation}
To diagonalize Hamiltonian (\ref{habrep2HO}), we introduce the squeezing operators
\begin{eqnarray}
S_{a}(r_{a})=e^{r_{a}({a}^{2}-{a}^{\dagger 2})/2},\hspace{0.5 cm}
S_{b}(r_{b})=e^{r_{b}({b}^{2}-{b}^{\dagger 2})/2},
\end{eqnarray}
where the two real squeezing parameters are defined by
\begin{eqnarray}
r_{a}=\frac{1}{2}\ln(\omega_{A}/\omega_{a}),\hspace{0.5 cm} r_{b}=\frac{1}{2}\ln(\omega_{B}/\omega_{b}).
\end{eqnarray}
The transformed Hamiltonian can be written as
\begin{equation}
H'={S}_{b}^{\dagger}(r_{b}){S}_{a}^{\dagger}(r_{a})UHU^{\dagger}{S}_{a}(r_{a}){S}_{b}(r_{b})
=\hbar\omega_{A}a^{\dagger}a+\hbar\omega_{B}b^{\dagger}b+\frac{1}{2}\hbar(\omega _{A}+\omega_{B}),
\end{equation}
where the unitary operator $U$ can be expressed with the operators $a$ and $b$ as
\begin{equation}
U=\exp\left[\frac{\theta}{2}\frac{\omega_{b}-\omega_{a}}{\sqrt{\omega_{a}\omega_{b}}}({a}^{\dagger }{b}^{\dagger }-{a}{b})-\frac{\theta}{2}\frac{\omega_{b}+\omega_{a}}{\sqrt{\omega_{a}\omega_{b}}}({a}^{\dagger}{b}-{a}{b}^{\dagger})\right].
\end{equation}
The eigenstates of the Hamiltonian $H'$ can be obtained as
\begin{eqnarray}
H'\vert m\rangle_{a}\vert n\rangle_{b}=E_{m,n}\vert m\rangle_{a}\vert n\rangle_{b},\hspace{0.3 cm}m,n=0,1,2,\cdots,
\end{eqnarray}
where the eigenvalues are defined in Eq.~(\ref{eigenvalueAB}).
The eigensystem of the Hamiltonian $H$ can be obtained as
\begin{eqnarray}
HU^{\dagger}{S}_{a}(r_{a}){S}_{b}(r_{b})\vert m\rangle_{a}\vert n\rangle_{b}=E_{m,n}U^{\dagger}{S}_{a}(r_{a}){S}_{b}(r_{b})\vert m\rangle_{a}\vert n\rangle_{b}.
\end{eqnarray}
As a result, the ground state of the system can be expressed as
\begin{eqnarray}
\vert G\rangle=U^{\dagger}{S}_{a}(r_{a}){S}_{b}(r_{b})\vert 0\rangle_{a}\vert 0\rangle_{b}.
\end{eqnarray}
In general, it is hard to write out the ground state in the number state representation. However, we can obtain a number-state expansion of the ground state in the degenerate two-oscillator case~\cite{Fan1992}, i.e., $\omega _{a}=\omega _{b}$. In this case, we have
$U=\exp[-(\pi/4)(a^{\dagger}b-ab^{\dagger})]$ and the ground state becomes
\begin{eqnarray}
\vert G\rangle=\exp\left[(\pi/4)({a}^{\dagger}{b}-{a}{b}^{\dagger})\right]{S}_{a}(r_{a}){S}_{b}(r_{b})|0\rangle_{a}|0\rangle_{b}.\label{Gs}
\end{eqnarray}
By expanding the squeezing operators, we then have
\begin{eqnarray}
\vert G\rangle&=&{U}^{\dagger}{S}_{a}(r_{a}){S}_{b}(r_{b})|0\rangle_{a}|0\rangle_{b}\nonumber\\
&=&\frac{1}{\sqrt{\cosh r_{a}}\sqrt{\cosh r_{b}}}\exp\left[\frac{\pi}{4}\left({a}^{\dagger}{b}-{a}{b}
^{\dagger}\right)\right]\exp\left( -\frac{1}{2}{a}^{\dagger 2}\tanh r_{a}\right)\exp
\left(-\frac{1}{2}{b}^{\dagger 2}\tanh r_{b}\right)|0\rangle_{a}|0\rangle_{b}.
\end{eqnarray}
In terms of the relations
\begin{subequations}
\begin{align}
\exp\left[\frac{\pi}{4}\left({a}^{\dagger}{b}-{a}{b}^{\dagger}\right)\right]
=&\exp\left({a}^{\dagger}{b}\right)\exp\left[\frac{1}{2}\left(\ln 2\right)\left( {a}^{\dagger}{a}-{b}^{\dagger}{b}\right)\right]\exp\left(-{a}{b}^{\dagger}\right),\\
\exp\left(-\frac{1}{2}{a}^{\dagger 2}\tanh r_{a}\right) |0\rangle _{a}
=&\sum_{m=0}^{\infty }\frac{\sqrt{\left( 2m\right)!}}{m!}\left(-\frac{1}{2}\tanh r_{a}\right) ^{m}|2m\rangle _{a},\\
\exp\left( -\frac{1}{2}{b}^{\dagger 2}\tanh r_{b}\right)|0\rangle _{b}
=&\sum_{n=0}^{\infty }\frac{\sqrt{\left(2n\right)!}}{n!}\left(-\frac{1}{2
}\tanh r_{b}\right)^{n}|2n\rangle_{b},
\end{align}
\end{subequations}
we then obtain
\begin{eqnarray}
\vert G\rangle&=&{U}^{\dagger}{S}_{a}\left(r_{a}\right){S}_{b}\left(
r_{b}\right)|0\rangle_{a}|0\rangle_{b}=\frac{1}{\sqrt{\cosh r_{a}}\sqrt{\cosh r_{b}}}\sum_{m,n=0}^{\infty}\frac{
\sqrt{\left(2m\right)!}\sqrt{\left(2n\right)!}}{m!n!}\left(-\frac{1}{2}\tanh r_{a}\right)^{m}\left(-\frac{1}{2}\tanh
r_{b}\right)^{n}\nonumber\\
&&\times\sum_{l=0}^{2m}\frac{\left(-1\right)^{l}}{l!}\sqrt{\frac{\left(
2m\right)!}{\left(2m-l\right)!}}\sqrt{\frac{\left(2n+l\right)!}{\left(
2n\right)!}}\exp\left[\left(m-n-l\right)\left(\ln 2\right)\right]\nonumber\\
&&\times\sum_{s=0}^{2n+l}\frac{1}{s!}\sqrt{\frac{\left(2m-l+s\right)!}{
\left(2m-l\right)!}}\sqrt{\frac{\left(2n+l\right)!}{\left(2n+l-s\right)
!}}|2m-l+s\rangle_{a}|2n+l-s\rangle_{b}.
\end{eqnarray}
It can be seen that the superposition components in the ground state are even parity states. This property can be confirmed because the transform $U$ conserves the excitation number and the squeezing operators change the excitation number two by two, without changing the parity.
%%%%%%%%%%%%%%%%%%%%%%%%%
\begin{figure}[tbp]
\includegraphics[width=0.7\textwidth]{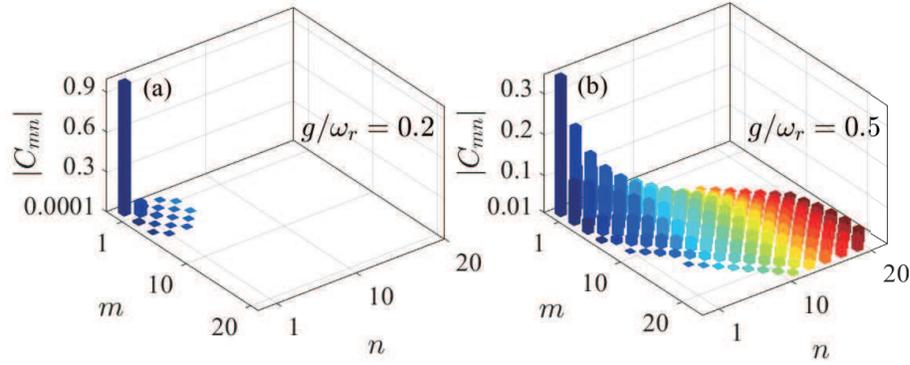}
\caption{The absolute values $\vert C_{m,n} \vert$ of the probability amplitudes for the ground state $G$ in the degenerate two-oscillator case when the coupling strength takes (a) $g/\omega_{r}=0.2$ and (b) $g/\omega_{r}=0.5$.}
\label{Cmnvalue}
\end{figure}
%%%%%%%%%%%%%%%%%%%%%%%%%

\textbf{Ground-state entanglement and quadrature squeezing.}  We study the ground-state entanglement in this system by calculating the logarithmic negativity. For the two-oscillator system, if the coupling is sufficiently weak, i.e., $g\ll\{\omega_{a},\omega_{b}\}$, the interaction Hamiltonian between the two oscillators can be reduced by the RWA as $H_{I}\approx g(a^{\dag}b+b^{\dag}a)$, which conserves the number of excitations. In this case, the ground state of the system is a trivial direct product of two vacuum states $\vert0\rangle_{a}\vert0\rangle_{b}$, which does not contain excitations. In the presence of the CR terms, the $\vert0\rangle_{a}\vert0\rangle_{b}$ is not an eigenstate of the system and the ground state will possess excitations. Below, we use numerical method to obtain the ground state of Hamiltonian~(\ref{eqHtwom}) and calculate the ground state entanglement between the two oscillators. In the presence of the CR terms, the ground state of the two-oscillator system can be expressed as
\begin{equation}
\vert G\rangle=\sum_{m,n=0}^{\infty}C_{m,n}\vert m\rangle_{a}\vert n\rangle_{b},\label{12ground}
\end{equation}
where these superposition coefficients are given by $C_{m,n}=\;_{a}\langle m\vert\;_{b}\!\langle n\vert G\rangle$, which should be solved numerically. The $\langle G\vert a^{\dag}a\vert G\rangle\neq0$ and $\langle G\vert b^{\dag}b\vert G \rangle\neq0$ reveal that the ground state of the system contains excitations. These excitations in the ground state are called virtual excitations because these excitations cannot be extracted from the system.

The effect of the virtual excitations can be seen from the probability amplitudes in the ground state. The distribution of these probability amplitudes can also exhibit the parity of the ground state. As the ground state is an even parity state, and hence these probability amplitudes associated with the odd parity basis states will disappear. In Fig.~\ref{Cmnvalue}, we show the absolute values of these probability amplitudes $\vert C_{m,n}\vert$. Here we can see that the values of $\vert C_{m,n}\vert$ decrease with the increase of $m$ and $n$ and that there is a symmetric relation $\vert C_{m,n}\vert=\vert C_{n,m}\vert$. In addition, the values of these odd-parity probability amplitudes $C_{m,n}$ with $m+n$ being an odd number are zero, which is a consequence of the fact that the ground state is an even-parity state.

We also calculate the average excitation numbers $\langle a^{\dag}a\rangle$ and $\langle b^{\dag}b\rangle$ in the ground state $\vert G\rangle$ as
\begin{equation}
\langle a^{\dag}a\rangle=\langle b^{\dag}b\rangle=\frac{1}{2}(\sinh^{2}r_{a}+\sinh^{2}r_{b}),
\end{equation}
where we have used the formula,
\begin{subequations}
\begin{align}
UaU^{\dag}=&(a+b)/\sqrt{2},\hspace{0.3 cm}UbU^{\dag}=(b-a)/\sqrt{2},\\
{S}_{a}^{\dagger}(r_{a})a{S}_{a}(r_{a})=&a\cosh r_{a}-a^{\dagger}\sinh r_{a},\hspace{0.3 cm}
{S}_{b}^{\dagger}(r_{b})b{S}_{b}(r_{b})=b\cosh r_{b}-b^{\dag}\sinh r_{b}.
\end{align}
\end{subequations}

\begin{figure}[tbp]
\includegraphics[width=0.7\textwidth]{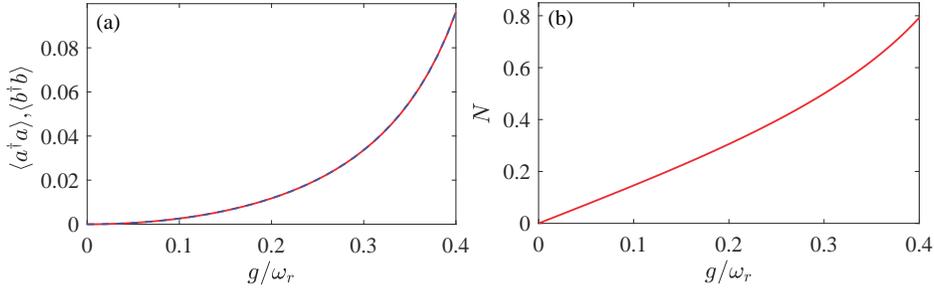}
\caption{(a) The average excitation numbers $\langle a^{\dag}a\rangle$, $\langle b^{\dag}b\rangle$ and (b) the logarithmic negativity in the ground state $\vert G\rangle$ of the degenerate two-oscillator system as functions of the ratio $g/\omega_{r}$.}
\label{groundstate}
\end{figure}

In Fig.~\ref{groundstate}a, we show the average excitation numbers $\langle a^{\dag}a\rangle$ and $\langle b^{\dag}b\rangle$ in the ground state $\vert G\rangle$ as functions of the scaled coupling strength $g/\omega_{r}$ in the degenerate oscillator case $\omega_{a}=\omega_{b}=\omega_{r}$. These results show that the average excitation numbers of the two modes are identical (two curves overlap each other). This is because the corotating terms conserve the excitations and the CR terms create simultaneously the excitations in the two modes. The average excitation numbers increase with the coupling strength since a larger coupling strength corresponds to a faster excitation creation.

The degree of entanglement between the two oscillators $a$ and $b$ in the ground state of the system can be obtained by calculating the logarithmic negativity. Combining with Eq.~(\ref{12ground}), the density matrix of the ground state can be written as
\begin{equation}
\rho=\sum_{m,n,j,k=0}^{\infty}C_{m,n}C_{j,k}^{\ast}\vert m\rangle_{a}\vert n\rangle_{b}\;_{a}\!\langle j\vert_{b}\!\langle k\vert.\label{groundsta2Rs}
\end{equation}
%%%%%%%%%%%%%%%%%%%%%%%%%
%%%%%%%%%%%%%%%%%%%%%%%%%
The degree of entanglement of the ground state can be quantized by calculating the logarithmic negativity~\cite{Vidal2002,Plenio2005}. For a bipartite system described by the density matrix $\rho$, the logarithmic negativity can be defined by
\begin{equation}
N=\log_{2}\left\Vert\rho^{T_{b}}\right\Vert_{1},\label{eqlog}
\end{equation}
where $T_{b}$ denotes the partial transpose of the density matrix $\rho$ of the system with respect to the oscillator $b$, and the trace norm $\Vert\rho^{T_{b}}\Vert_{1}$ is defined by
\begin{equation}
\left\Vert\rho^{T_{b}}\right\Vert_{1}=\text{Tr}\left[\sqrt{(\rho^{T_{b}})^{\dag}\rho^{T_{b}}}\right]. \label{tbrou}
\end{equation}

Using Eqs.~(\ref{groundsta2Rs}),~(\ref{eqlog}), and~(\ref{tbrou}), the logarithmic negativity of ground state of the two coupled oscillators can be obtained. In Fig.~\ref{groundstate}b, we show the logarithmic negativity $N$ as a function of the coupling parameter $g/\omega_{r}$. The curve shows that the degree of entanglement between the two oscillators in the ground state monotonically increases over the entire range of $g$. This is because the CR terms in Hamiltonian~(\ref{eqHtwom}) cause the virtual excitations in the ground state of the system and maintain the quantum entanglement between the two oscillators. If the CR terms are discarded, then the ground state of the system becomes a separate state $\vert 0\rangle_{a}\vert 0\rangle_{b}$.

We also study the quadrature squeezing in the ground state by calculating the fluctuations of the rotated quadrature operators. We introduce the rotated quadrature operators for the two modes as
\begin{equation}
X_{o}(\theta_{o})=(o^{\dag}e^{i\theta_{o}}+oe^{-i\theta_{o}}),\hspace{0.3 cm} X_{o}(\theta_{o}+\pi/2)=i(o^{\dag}e^{i\theta_{o}}-oe^{-i\theta_{o}}),\hspace{0.3 cm} o=a,b.
\end{equation}
The commutation relation of the above two rotated quadrature operators is
\begin{equation}
\left[X_{o}(\theta_{o}),X_{o}(\theta_{o}+\pi/2)\right]=2i,\hspace{0.3 cm} o=a,b.\label{commutation}
\end{equation}
According to the uncertainty relation, we have
\begin{equation}
\Delta X_{o}^{2}(\theta_{o})\Delta X_{o}^{2}(\theta_{o}+\pi/2)\geq 1,\hspace{0.3 cm}o=a,b.
\end{equation}
Then the quadrature squeezing appears along the angle $\theta_{o}$ if the variances of the rotated quadrature operators satisfy the relation\cite{Scully1997}
\begin{equation}
\Delta X_{o}^{2}(\theta_{o})<1,\hspace{0.3 cm}o=a,b.
\end{equation}
For the ground state $\vert G\rangle$ given in Eq.~(\ref{Gs}), the variances of the rotated quadrature operators can be obtained as
\begin{eqnarray}
\Delta X_{a}^{2}\left( \theta _{a}\right)  &=&\left\langle G\right\vert
X_{a}^{2}\left( \theta _{a}\right) \left\vert G\right\rangle -\left(
\left\langle G\right\vert X_{a}\left( \theta _{a}\right) \left\vert
G\right\rangle \right) ^{2}  \nonumber \\&=&-\cos \left( 2\theta _{a}\right) \left( \sinh r_{a}\cosh r_{a}+\sinh r_{b}\cosh r_{b}\right) +\sinh ^{2}r_{a}+\sinh ^{2}r_{b}+1.\label{variances}
\end{eqnarray}
When we exchange the subscripts $a$ and $b$ in Eq.~(\ref{variances}), the expression does not change for a given rotating angle. This means that, in the resonance case $\omega_{a}=\omega_{b}=\omega_{r}$, the squeezing is the same for the two bosonic modes in the ground state. This point can also be seen from Hamiltonian~(\ref{eqHtwom}), which is symmetric under the exchange of the subscripts and operators for the two modes in the resonance case.
	
\begin{figure}[tbp]
\includegraphics[width=0.7\textwidth]{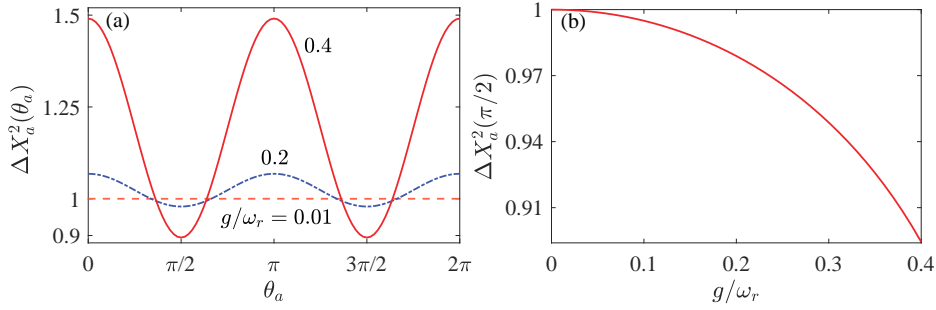}
\caption{(a) The variance $\Delta X_{a}^{2}(\theta_{a})$ of the rotated quadrature operators as a function of the angle $\theta_{a}$ in the resonant case $\omega_{a}=\omega_{b}=\omega_{r}$ when $g/\omega_{r}=0.01$, $0.2$, and $0.4$. (b) The variance $\Delta X_{a}^{2}(\pi/2)$ as a function of the coupling strength $g/\omega_{r}$ in the resonant case $\omega_{a}=\omega_{b}=\omega_{r}$.}
\label{squeezed}
\end{figure}

In Fig.~\ref{squeezed}a, we show the variance $\Delta X_{a}^{2}(\theta_{a})$ as a function of the rotating angle $\theta_{a}$ in the resonance case $\omega_{a}=\omega_{b}=\omega_{r}$. Here we can see that the variance  $\Delta X_{a}^{2}(\theta_{a})$ is periodic function of $\theta_{a}$ and that the minimum of $\Delta X_{a}^{2}(\theta_{a})$ is obtained at $\theta_{a}=\pi/2$ and $3\pi/2$. Note that in the present case $( \sinh r_{a}\cosh r_{a}+\sinh r_{b}\cosh r_{b}) < 0$. We also show the variance $\Delta X_{a}^{2}(\pi/2)$ as a function of the coupling strength $g/\omega_{r}$ in the resonant case $\omega_{a}=\omega_{b}=\omega_{r}$, as shown in Fig.~\ref{squeezed}b. We observe that the squeezing increases with the scaled coupling strength $g/\omega_{r}$. This is because the quadrature squeezing is caused by the CR interaction terms.
%%%%%%%%%%%%%%%%%%%%%%%%%
%%%%%%%%%%%%%%%%%%%%%%%%%

\textbf{Dynamics of quantum entanglement.}  The phenomenon of quantum entanglement accompanied with virtual excitations can also be seen by analyzing the entanglement dynamics of the system. We consider the case in which the system is initially in the zero-excitation state $\vert 0\rangle_{a}\vert 0\rangle_{b}$. In the closed-system case, a general state of the system can be written as
\begin{equation}
\vert\psi(t)\rangle=\sum_{m,n=0}^{\infty}A_{m,n}(t)\vert m\rangle_{a}\vert n\rangle_{b}.\label{eqHamplitude}
\end{equation}
By substituting Eqs.~(\ref{eqHtwom}) and (\ref{eqHamplitude}) into the Schr\"{o}dinger equation, the equations of motion for these probability amplitudes $A_{m,n}(t)$ are obtained as
\begin{eqnarray}
\dot{A}_{m,n}(t)=&-i(\omega_{a}m+\omega_{b}n)A_{m,n}(t)
-ig\left[\sqrt{(m+1)(n+1)}A_{m+1,n+1}(t)+\sqrt{mn}A_{m-1,n-1}(t)\right.\nonumber\\
&\left.+\sqrt{m(n+1)}A_{m-1,n+1}(t)+\sqrt{(m+1)n}A_{m+1,n-1}(t)\right].
\label{ampamn}
\end{eqnarray}
For the initial state $\vert 0\rangle_{a}\vert 0\rangle_{b}$, the initial condition of these probability amplitudes reads $A_{m,n}(0)=\delta_{m,0}\delta_{n,0}$. By numerically solving Eq.~(\ref{ampamn}) under this initial condition, the evolution of these probability amplitudes can be obtained. Using Eqs.~(\ref{eqlog}), (\ref{tbrou}), and (\ref{eqHamplitude}), we can calculate numerically the average excitation numbers $\langle a^{\dag}a\rangle$ and $\langle b^{\dag}b\rangle$ and the logarithmic negativity of the state $\vert\psi(t)\rangle$.

\begin{figure}[tbp]
\includegraphics[width=0.7\textwidth]{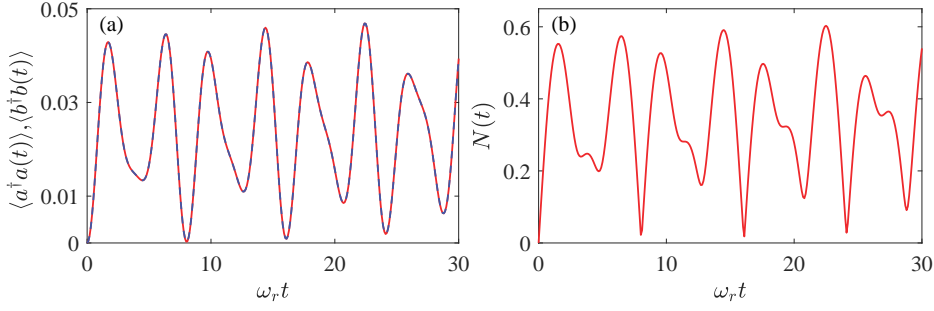}
\caption{Dynamics of (a) the average excitation numbers $\langle a^{\dag}a(t)\rangle$, $\langle b^{\dag}b(t)\rangle$ and (b) the logarithmic negativity when the degenerate two-oscillator system is initially in the state $\vert 0\rangle_{a}\vert 0\rangle_{b}$. The used parameter is $g/\omega_{r}=0.2$.}
\label{closedsystem}
\end{figure}

In Fig.~\ref{closedsystem}a, we show the time evolution of the average excitation numbers $\langle a^{\dag}a\rangle$ and $\langle b^{\dag}b\rangle$ in modes $a$ and $b$. Here we can see that, similar to the ground state case, the average excitation numbers in the two modes are identical (the two curves overlap each other). In addition, the average excitation numbers experience a periodic oscillation. In Fig.~\ref{closedsystem}b, we show the time dependence of the logarithmic negativity $N(t)$ of the state $\vert\psi(t)\rangle$. The curve shows that logarithmic negativity between the two oscillators also experiences a periodic oscillation. Here we choose the initial state of the system as $\vert 0\rangle_{a}\vert 0\rangle_{b}$, the existence of the CR terms still causes the appearance of virtual excitations, which leads to entanglement between the two oscillators. This result is different from that in the RWA case in which the CR terms are discarded in the two oscillators under the same initial state. When we discard the CR terms and choose the initial state as $\vert 0\rangle_{a}\vert 0\rangle_{b}$, which is the eigenstate of the corotating interaction term $g(a^{\dag}b+b^{\dag}a)$, the system will stay this state. Then there are no virtual excitations in the system and no quantum entanglement between the two oscillators.
%%%%%%%%%%%%%%%%%%%%%%%%%
%%%%%%%%%%%%%%%%%%%%%%%%%

We also study the influence of the environment dissipations on the dynamics of the system. As we consider the ultrastrong-coupling regime of the coupled system, we derive the quantum master equation in the normal-mode representation of these two coupled oscillators. We employ the standard Born-Markov approximation under the condition of weak system-bath couplings and short bath correlation times to derive the quantum master equation. The secular approximation is made by discarding these high-frequency oscillating terms including $\exp(\pm i\omega_{A}t)$, $\exp(\pm i\omega _{B}t)$, and $\exp[\pm i(\omega_{A}\pm\omega_{B})t]$. The quantum master equation in the normal-mode representation of Hamiltonian~(\ref{eqHtwom}) can be written as
\begin{equation}
\frac{d}{dt}\tilde{\rho}_{s}(t) =i[ \tilde{\rho}_{s}\left(
t\right) ,\tilde{H}] +\alpha _{1}\mathcal{D}\left[ A\right] \tilde{\rho}_{s}(t) +\alpha _{2}\mathcal{D}\left[ B\right] \tilde{\rho}_{s}\left( t\right),\label{eqrou12}
\end{equation}
with the effective decay rates
\begin{equation}
\alpha_{1}=(f_{1}+f_{2})^{2}\gamma_{a}+(f_{5}+f_{6})^{2}\gamma_{b},\hspace{0.3 cm}
\alpha_{2}=(f_{3}+f_{4})^{2}\gamma_{a}+(f_{7}+f_{8})^{2}\gamma_{b}.
\end{equation}
In Eq.~(\ref{eqrou12}), the dissipator
\begin{equation}
\mathcal D[o]\tilde{\rho}_{s}(t)=o\tilde{\rho}_{s}(t) o^{\dag}-[o^{\dag}o\tilde{\rho}_{s}(t)+\tilde{\rho}_{s}(t)o^{\dag}o]/2\label{superoperator}
\end{equation}
is the standard Lindblad superoperator that describes the dampings of the oscillators. The parameters $\gamma_{a}$ and $\gamma_{b}$ are the decay rates relating to the heat bath in contact with the oscillators $a$ and $b$, respectively. Here we consider the zero temperature environments such that the thermal excitation effect can be excluded.

In our simulations, we need to calculate the evolution of the system in the bare-mode representation, i.e., $\{a, b\}$. The relationship between the density matrix $\tilde{\rho}_{s}(t)$ in the normal-mode representation ( $\{A,A^{\dag},B,B^{\dag}\}$) and the density matrix $\rho_{s}(t)$ in the bare-mode representation ($\{a,a^{\dag},b,b^{\dag}\}$) is determined by the transformation $\tilde{\rho}_{s}(t)=U\rho_{s}(t)U^{\dag}$. Combining with Eq.~(\ref{Htildexp}), the quantum master equation in the bare-mode representation can be obtained as
\begin{equation}
\dot{\rho}_{s}(t)=i[\rho_{s}(t),H]+\alpha _{1}U^{\dag }\mathcal{D}[A]\tilde{\rho}_{s}(t) U+\alpha _{2}U^{\dag }\mathcal{D}[B] \tilde{\rho}
_{s}(t) U.\label{abmeq}
\end{equation}
The transformation relations between the operators $\{A,A^{\dag},B,B^{\dag}\}$ and $\{a,a^{\dag},b,b^{\dag}\}$ are
\begin{equation}
U^{\dag }AU=F_{1}a+F_{2}a^{\dag }-F_{3}b-F_{4}b^{\dag },\hspace{0.3 cm}U^{\dag}BU=F_{5}a+F_{6}a^{\dag }+F_{7}b+F_{8}b^{\dag },\label{ABoperators}
\end{equation}
where the forms of these coefficients $F_{i}$ ($i = 1,2,\cdots ,7,8$) are given by
\begin{subequations}
\label{Fi}
\begin{align}
F_{1,2} =&\frac{1}{2}\frac{\cos \left( \theta \right) }{\sqrt{\omega_{a}\omega _{A}}}\left( \omega _{A}\pm\omega _{a}\right),
\hspace{0.5 cm}F_{3,4}=\frac{1}{2}\frac{\sin \left( \theta \right) }{\sqrt{\omega _{b}\omega _{A}}}\left(\omega _{A}\pm\omega _{b}\right),\\
F_{5,6} =&\frac{1}{2}\frac{\sin \left( \theta \right) }{\sqrt{\omega_{a}\omega _{B}}}\left( \omega _{B}\pm\omega _{a}\right),
\hspace{0.5 cm}F_{7,8}=\frac{1}{2}\frac{\cos \left( \theta \right) }{\sqrt{\omega _{b}\omega _{B}}}\left(\omega _{B}\pm\omega _{b}\right).
\end{align}
\end{subequations}
By substituting Eqs.~(\ref{superoperator}),~(\ref{ABoperators}), and~(\ref{Fi}) into Eq.~(\ref{abmeq}), we obtain
\begin{eqnarray}
&&\alpha_{1}U^{\dag}\mathcal{D}[A]\tilde{\rho}_{s}(t)U+\alpha_{2}U^{\dag}\mathcal{D}[B]\tilde{\rho}_{s}(t)U\nonumber\\
&&=\beta _{1}\mathcal{D}\left[ a\right] \rho _{s}\left( t\right) +\beta _{2}
\mathcal{D}\left[ b\right] \rho _{s}\left( t\right) +\beta _{3}\mathcal{D}
[ a^{\dag }] \rho _{s}\left( t\right) +\beta _{4}\mathcal{D}[b^{\dag }] \rho _{s}\left( t\right)  \nonumber\\
&&+\beta _{5}\left(\mathcal{S}\left[ a,a\right] \rho _{s}\left( t\right) +
\mathcal{S}[ a^{\dag },a^{\dag }] \rho _{s}\left( t\right)
\right) +\beta _{6}\left( \mathcal{S}\left[ b,b\right] \rho _{s}\left(
t\right) +\mathcal{S}[ b^{\dag },b^{\dag }] \rho _{s}\left(
t\right) \right)  \nonumber\\
&&+\beta _{7}\left( \mathcal{S}[ a,b^{\dag }] \rho _{s}\left(
t\right) +\mathcal{S}[ b,a^{\dag }] \rho _{s}\left( t\right)
\right) +\beta _{8}\left( \mathcal{S}\left[ a,b\right] \rho _{s}\left(
t\right) +\mathcal{S}[ b^{\dag },a^{\dag }] \rho _{s}\left(
t\right) \right) \nonumber\\
&&+\beta _{9}\left( \mathcal{S}[ a^{\dag },b^{\dag }] \rho
_{s}\left( t\right) +\mathcal{S}\left[ b,a\right] \rho _{s}\left( t\right)
\right) +\beta _{10}\left( \mathcal{S}[ a^{\dag },b] \rho
_{s}\left( t\right) +\mathcal{S}[ b^{\dag },a] \rho _{s}\left(
t\right) \right),\label{dissipatorcorss}
\end{eqnarray}
where we introduce the superoperator as
\begin{equation}
\mathcal{S}[ o,o^{\prime}]\rho_{s}(t) =o \rho_{s}(t) o^{\prime}-[o^{\prime}o \rho_{s}(t) + \rho_{s}(t) o^{\prime}o]/2.
\end{equation}
The coefficients introduced in Eq.~(\ref{dissipatorcorss}) are defined by
\begin{subequations}
\begin{align}
\beta _{1} =&\alpha _{1}F_{1}^{2}+\alpha _{2}F_{5}^{2},\hspace{0.5 cm}\beta _{2}=\alpha_{1}F_{2}^{2}+\alpha _{2}F_{6}^{2}, \\
\beta _{3} =&\alpha _{1}F_{3}^{2}+\alpha _{2}F_{7}^{2},\hspace{0.5 cm}\beta _{4}=\alpha_{1}F_{4}^{2}+\alpha _{2}F_{8}^{2}, \\
\beta _{5} =&\alpha _{1}F_{1}F_{2}+\alpha _{2}F_{5}F_{6},\hspace{0.5 cm}\beta_{6}=\alpha _{1}F_{3}F_{4}+\alpha _{2}F_{7}F_{8}, \\
\beta _{7} =&\alpha _{2}F_{5}F_{7}-\alpha _{1}F_{1}F_{3},\hspace{0.5 cm}\beta_{8}=\alpha _{2}F_{5}F_{8}-\alpha _{1}F_{1}F_{4}, \\
\beta _{9} =&\alpha _{2}F_{6}F_{7}-\alpha _{1}F_{2}F_{3},\hspace{0.5 cm}\beta_{10}=\alpha _{2}F_{6}F_{8}-\alpha _{1}F_{2}F_{4},
\end{align}
\end{subequations}
Note that the cross terms between the two modes $a$ and $b$ in Eq.~(\ref{dissipatorcorss}) are induced by the interaction between the two oscillators.
%%%%%%%%%%%%%%%%%%%%%%%%%
\begin{figure}[tbp]
\includegraphics[width=0.7\textwidth]{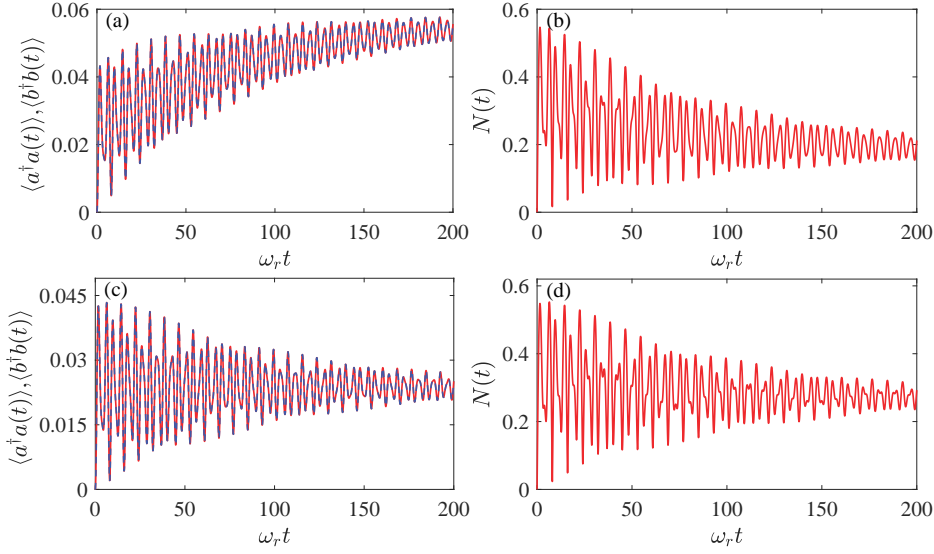}
\caption{Dynamics of (a), (c) the average excitation numbers $\langle a^{\dag}a(t)\rangle$, $\langle b^{\dag}b(t)\rangle$ and (b), (d) the logarithmic negativity as functions of the evolution time $t$ when the system is initially in the state $\vert0\rangle_{a}\vert0\rangle_{b}$. The parameters used are $g/\omega_{r}=0.2$ and $\gamma_{a}/\omega_{r}=\gamma_{b}/\omega_{r}=10^{-2}$. The results in panels (a) and (b) [(c) and (d)] are calculated with the microscopic quantum master equation (the phenomenological quantum master equation).}
\label{opensystem}
\end{figure}
%%%%%%%%%%%%%%%%%%%%%%%%%
For below calculations, we express the density matrix of the two-oscillator system in the bare-mode representation as
\begin{equation}
\rho_{s}(t)=\sum_{m,n,j,k=0}^{\infty}\rho_{m,n,j,k}(t)\vert m\rangle_{a}\vert n\rangle_{b}\;_{a}\!\langle j\vert_{b}\!\langle k\vert,\label{eqrou}
\end{equation}
with the density matrix elements $\rho_{m,n,j,k}(t)=\;_{a}\!\langle m\vert_{b}\!\langle n\vert\rho_{s}(t)\vert j\rangle_{a}\vert k\rangle_{b}$. For an initial state $\vert0\rangle_{a}\vert0\rangle_{b}$, the nonzero density matrix element is $\rho_{0,0,0,0}(0)=1$. By numerically solving Eq.~(\ref{abmeq}) under the initial condition, the time evolution of the density matrix $\rho_{s}(t)$ can be obtained.

Below we study the dynamics of the average excitation numbers and quantum entanglement in this system.
Based on Eq.~(\ref{abmeq}), the expressions of the average excitation numbers $\langle a^{\dag}a(t)\rangle$ and $\langle b^{\dag}b(t)\rangle$ can be expanded as
\begin{equation}
\langle a^{\dag}a(t)\rangle=\text{Tr}[a^{\dag}a\rho_{s}(t)]=\sum_{m,n=0}^{\infty}m\rho_{m,n,m,n}(t),\hspace{0.3 cm}
\langle b^{\dag}b(t)\rangle=\text{Tr}[b^{\dag}b\rho_{s}(t)]=\sum_{m,n=0}^{\infty}n\rho_{m,n,m,n}(t).
\end{equation}
Therefore, the average excitation numbers $\langle a^{\dag}a(t)\rangle$ and $\langle b^{\dag}b(t)\rangle$ can be obtained by solving the equations of motion for these density matrix elements in the number-state representation.

In Fig.~\ref{opensystem}a, the dynamics of the average excitation numbers $\langle a^{\dag}a(t)\rangle$ and $\langle b^{\dag}b(t)\rangle$ is shown in the open-system case with different time $t$. We observe that the two excitation numbers $\langle a^{\dag}a(t)\rangle$ and $\langle b^{\dag}b(t)\rangle$ overlap each other and initially experience a large oscillation. With the increase of time $t$, the oscillation amplitudes of the average excitation numbers decrease gradually. In the long-time limit $t\gg 1/\gamma_{a,b}$, the average excitation numbers will reach steady values due to the dissipations.

The entanglement of the density matrix $\rho_{s}(t)$ can be quantified by calculating the logarithmic negativity. In terms of Eqs.~(\ref{eqlog}), (\ref{abmeq}), and (\ref{eqrou}), the logarithmic negativity of the state $\rho_{s}(t)$ can be obtained numerically. In Fig.~\ref{opensystem}b, we show the logarithmic negativity $N(t)$ of the density matrix $\rho_{s}(t)$ versus the time $t$. The result shows that the logarithmic negativity oscillates very fast due to the free evolution of the system. We also find that the envelope of the logarithmic negativity converges gradually with the evolution time $t$ and eventually reaches a stable value due to the dissipations. The time scale of the oscillation-pattern decay for the logarithmic negativity is very similar to that of the excitations created by the CR interaction terms. In particular, we find that there exists steady-state entanglement due to the presence of the CR interaction terms in this system.

In this work, we consider the ultrastrong-coupling regime and hence the quantum master equation is derived in the normal mode representation.  For comparison, we show in Figs.~\ref{opensystem}c,d the evolution of the average excitation numbers and the logarithmic negativity calculated by solving the phenomenological quantum master equation, which is obtained by adding the dissipators of two free bosonic modes into the Liouville equation,
\begin{equation}
\dot{\rho}_{s}(t)=i[{\rho}_{s}(t),{H}]+\gamma_{a}\mathcal{D}[a]{\rho}_{s}(t)
+\gamma_{b}\mathcal{D}[b]{\rho}_{s}(t).\label{phenquantum}
\end{equation}
The initial state of the system is the same as that considered in the microscopic quantum master equation. We see from Fig.~\ref{opensystem} that, for the average excitation numbers, though these results can approach steady-state values, the envelop and the oscillation amplitude are different for the results obtained with two different quantum master equations. However, for the logarithmic negativity, we find that the difference between the two results exists but is small when $g/\omega_{r}=0.2$. We checked the fact that the difference will increase as the increase of the ratio $g/\omega_{r}$. Therefore, the microscopic quantum master equation should be used in the ultrastrongly-coupled-oscillator system.

\section*{Conclusion \label{Conclusion}}

In conclusion, we have studied quantum entanglement in an ultrastrongly-coupled two-harmonic-oscillator system. Concretely, we have studied the ground-state entanglement by calculating the logarithmic negativity of the ground state. Here, the quantum entanglement is maintained by the virtual excitations generated by the CR terms and bounded in the ground state. We have also studied the dynamics of quantum entanglement of the system. By microscopically deriving a quantum master equation in the normal-mode representation of the two oscillators, we analyzed the influence of the dissipations on the entanglement dynamics and found that there exists steady-state entanglement in this system.

\section*{Acknowledgements}
J.-F.H. is supported in part by the National Natural Science Foundation of China (Grant No. 11505055), Scientific Research Fund of Hunan Provincial Education Department (Grant No. 18A007), and Natural Science Foundation of Hunan Province, China (Grant No. 2020JJ5345). J.-Q.L. is supported in part by National Natural Science Foundation of China (Grants No.~11822501, No.~11774087, and No.~11935006), Natural Science Foundation of Hunan Province, China (Grant No.~2017JJ1021), and Hunan Science and Technology Plan Project (Grant No.~2017XK2018).

\section*{Author contributions}
J.-Q.L. provided the idea, contributed to the analyses of theory and data, and wrote the manuscript. J.-F.H. contributed to the analyses of theory and data, and modified the manuscript. J.-Y.Z. performed the numerical calculations, and contributed to the interpretation of the numerical results and the writing of the manuscript. Y.-H.Z. contributed to some theoretical calculations and checked the results. X.-L.Y. checked  the theoretical calculations. All authors reviewed the manuscript.

\section*{Additional information}
\textbf{Competing interests}
: The authors declare no competing interests.

\end{document}